\documentclass{article}
\usepackage{graphicx}
\newcommand{\bfr}{\begin{flushright}}
\newcommand{\efr}{\end{flushright}}
 
\begin{document}
\title{Vacuum energy for Yang-Mills fields in $R^d\times S^1$: One-loop,
two-loop, and beyond}
\author{Kiyoshi Shiraishi\\
and\\
Satoru Hirenzaki}
\date{Zeitschrift f\"ur Physik {\bf C53} (1992) pp. 91--96}
\maketitle
\begin{abstract}
The vacuum energy is calculated for Yang-Mills (YM) system defined in
$D$ dimensional space-time of $S^1\times R^d$ ($D=d+1$), where the
possibility of the YM fields to acquire the vacuum expectation values
on $S^1$ is taken into account. The vacuum energy has already been
obtained to the order of one-loop in many people. Here we calculate the
vacuum energy in $D$ dimensions to two-loop order. With an intention to
reach higher loops, an approximation method is proposed, which is
especially effective in higher dimensions. By this method, we can treat
the higher-loop contributions of YM interactions as easily as we treat
one-loop effect. As a check, we show reproduction of the two-loop
contribution ($D$-dependence of the coefficient as well as the
functional form) when the coupling constant is small. This
approximation method is useful not only for the Kaluza-Klein theories
but also for the finite temperature-density system (as a quark-gluon
plasma).
\end{abstract}

\section{Introduction}
The quantum field theories in the space-time of non-trivial
topology have been concerned with various physical situations. For
instance, the Casimir effect \cite{1} is a well-known example of them.
This effect is shown experimentally by measuring the force between the
conducting plates. Thus it is important to investigate the quantum
effect to understand the forces working in the topologically
non-trivial space-time and their integrals, that is, the ``vacuum
energy'' (see, however, \cite{2}).

The vacuum energy and symmetry breaking by scalar
fields have been studied in the compact spaces including
a torus and a sphere \cite{3}. The quantum effects in compact
spaces being as internal ones have been considered in
relation to Kaluza-Klein theories \cite{4,5}. It has been
pointed out that the quantum effect in a topologically
non-trivial space plays an important role in symmetry
breaking in string theory \cite{6} and in Kaluza-Klein
theories \cite{7}.

Now we take up as an example the quantum effect
in Kaluza-Klein theories \cite{5,7} in order to see how the
calculation of the effect has been performed. In generic
Kaluza-Klein theories the stability of the extra space has
been considered, where the estimation of the strength of
Casimir-like force plays the crucial role. The contribution
of gravitons and matter fields to the vacuum energy has
been also computed \cite{5}.

On the other hand, the model with symmetry breaking by gauge fields defined on the extra space has been
investigated. The vacuum energy in the presence of the
background gauge field has been calculated and is used
to determine the true vacuum among vacua which belongs
to various gauge symmetry \cite{7}.

The approaches so far are to calculate the vacuum
energies to one-loop order. They are expressed symbolically as
\begin{equation}
(\mbox{one-loop vac. energy}) = \log {\rm Det~} (-\Delta)\,,
\end{equation}
where $A$ stands for the generalized Laplacian associated
with the field under consideration. That is the inverse of
propagator.

The reasons why the one-loop approximation is often
examined are: i) the technique of calculation and regularization methods (for instance, the if-function regularization
and the dimensional regularization) are well-defined and
can be generalized to an arbitrary dimensional case, and
ii) more over we can carry out one-loop calculation only
by the knowledge of the free propagator and thus we
need not to worry about the renormalization scheme (of
interactions) in the way of calculation.

However we know how important the interaction of
fields is in the quantum effect. It is true that the interaction
has an essential effect on the quantum effect in the gauge
theory including QCD in four dimensions (see \cite{8} in the
context of Kaluza-Klein theory).

In the next section we perform the two-loop calculation of the vacuum
energy for YM fields on the $R^d\times S^1$ background topology. This is
a generalization of the two-loop calculation of the free energy for YM
fields at finite temperature which has been calculated before.
For later use, the background gauge field is taken into
account in our calculation. In the present paper, we discuss
only the vacuum energy (the free energy) for simplicity.

The two-loop calculations of the vacuum energy including Dirac fermions are also shown in Sect. 2. These
have not been sufficiently surveyed even in the four-dimensional
finite-temperature system. We consider the fermions in fundamental and
adjoint representations, possessing a general boundary condition, an
arbitrary dimension in the presence of a background gauge field
on $S^1$.

In Sect. 3, we try to include the effect YM interactions
beyond two-loops. The approximation scheme proposed
there is valid for a large number of dimensions, $D$.

The last section is devoted to the summary and consideration of future
problems.

Before closing this section, we mention some comments. In the
background configuration considered in this paper, both the curvature
and the field strength of gauge field vanish. Thus we naively expect
vanishing expectation value for local counter terms. That is why
there is no difficulty in obtaining the vaccum energy in
the general dimensions. Though we can proceed to a neat
discussion on the regularization making use of the
operator regularization method \cite{9}, that is beyond the
scope of this paper.

\section{The two-loop contributions to vaccum energy
of Yang-Mills system with fermions in $R^d\times S^1$ space}
In finite-temperature systems, or systems in Euclidean
space-time ($R^3\times S^1$), vacuum energy of YM fields to
two-loop order has already been calculated by many
authors \cite{10,11}. We compute the vacuum energy in an
arbitrary space-time dimensions $D(=d+1)$ in this
section. In the present paper, we restrict ourselves to the
case for $SU(2)$ gauge group throughout this paper.

We assume $R^d\times S^1$ as the background geometry. The
circumference of $S^1$ is set to $L$. We take gauge field
condensation on $S^1$ into consideration. It is parametrized
as the following (matrix-)form.
\begin{equation}
\langle A_I\rangle=\frac{\phi}{2gL}\left(
\begin{array}{cc}
1 & 0\\
0 & -1
\end{array}
\right)\,, 
\label{2.1}
\end{equation}
where $g$ is the YM coupling constant. Because of the
presence of the background field, we use the so-called
covariant background gauge method in the calculations,
in which quantum and classical fields are neatly separated (see, for example,
\cite{12}). Furthermore, we take the Feynman gauge $(\xi=1)$ unless we
indicate the gauge explicitly.

\begin{figure}[ht]
\begin{center}
\includegraphics[width=3cm,clip]{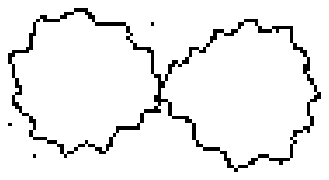}
\includegraphics[width=3cm,clip]{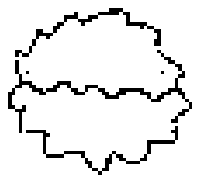}
\includegraphics[width=3cm,clip]{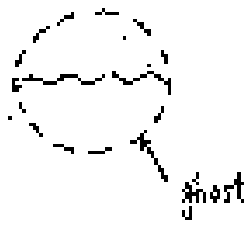}\\
($a$) \hspace{3cm} ($b$) \hspace{3cm}($c$)
\caption{The two-loop contributions of YM bosons to the vacuum
energy
}
\label{f1}
\end{center}
\end{figure}

There are three graphs for two-loop vaccum diagrams
of YM fields, of the order of $g^2$ (see Fig.~1). Each
contribution to the vacuum energy in $R^d\times S^1$ space is
\begin{eqnarray}
(a):& & +\frac{1}{2}g^2 D(D-1)I(\phi)\{I(\phi)+2I(0)\}\,,\label{2.2a}\\
(b):& & -\frac{3}{2}g^2 (D-1)I(\phi)\{I(\phi)+2I(0)\}\,,\label{2.2b}\\
(c):& & +\frac{1}{2}g^2I(\phi)\{I(\phi)+2I(0)\}\,,\label{2.2c}
\end{eqnarray}
where the function $I$ is defined as
\begin{equation}
I(x)=\frac{\Gamma\left(\frac{D-2}{2}\right)}{2\phi^{D/2}L^{D-2}}
\sum_{k=1}^\infty\frac{\cos kx}{k^{D-2}}\,.
\label{2.3}
\end{equation}
$I(x)$ comes from the loop integration obtained after
performing a suitable regularization which is chosen as
in the four-dimensional case.

Summing over the three contributions, we obtain the
two-loop contribution to vacuum energy for pure YM
system:
\begin{equation}
(2-\mbox{log vac. energy for YM})
=+\frac{1}{2}g^2(D-2)^2 I(\phi)\{I(\phi)+2I(0)\}\,.
\label{2.4}
\end{equation}
This calculation is a generalization of the result of \cite{11}.
Here, we must note: i) the functional form of each
contribution of diagram (a-c) is identical, ii) however,
the dependence of coefficient on the dimensionality $D$ is
different from each other, iii) if $D=2$, the total contribution
vanishes (as expected).

Next we study the fermion loops. The one-loop contribution of the fermion to vacuum energy can be easily
obtained. We show here the result for the two-loop
contribution which is made of fermion and YM boson
in a general dimension $D$. This is also of the order of $g^2$.

In a finite-temperature system, the calculation for fermion
contribution is important particularly in the study of the so-called
quark-gluon plasma. For future applications, we consider a generic
boundary condition of fermions on $S^1$ as well as a background gauge
field on $S^1$. We assume that the fermion fields has the following
dependence on the translation in the coordinate of $S^1$, which we
denote $y$: 
\begin{equation}
\Psi(y+L) = e^{i\delta}\Psi(y)\,.
\label{2.5}
\end{equation}
The only graph to calculate is given in Fig.~2.

\begin{figure}[ht]
\begin{center}
\includegraphics[width=2cm]{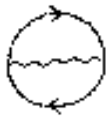}
\caption{The two-loop contributions of fermion and YM boson to
the vacuum energy
}
\label{f2}
\end{center}
\end{figure}

We consider Dirac fermions which belong to fundamental and adjoint
representations. After a small handwork, we obtain:
\begin{eqnarray}
& &\mbox{(2-loop vac. energy for fund. fermion)}\nonumber \\
&=&-\frac{1}{2}g^22^{[D/2]}\frac{D-2}{2}\nonumber \\
&
&\cdot\left[\frac{1}{2}\left\{I\left(\delta+\frac{\phi}{2}\right)+
I\left(\delta-\frac{\phi}{2}\right)\right\}\{I(0)+2I(\phi)\}\right.
\nonumber \\
&
&-I\left(\delta+\frac{\phi}{2}\right)I\left(\delta-\frac{\phi}{2}\right)
\nonumber \\
&
&\left.-\frac{1}{4}\left\{I\left(\delta+\frac{\phi}{2}\right)^2+
I\left(\delta-\frac{\phi}{2}\right)^2\right\}\right]
\label{2.6a}\\
& &\mbox{(2-loop vac. energy for adj. fermion)}\nonumber \\
&=&-\frac{1}{2}g^22^{[D/2]}\frac{D-2}{2}\nonumber \\
& &\cdot\{2\{I(\delta+\phi)+I(\delta-\phi)\}\{I(0)+I(\phi)-I(\delta)\}
\nonumber \\
& &+4I(\phi)I(\delta)-I(\delta+\phi)^2-I(\delta-\phi)^2\}\,,
\label{2.6b}
\end{eqnarray}
where $[~]$ is Gauss' symbol.

These results for general conditions are obtained for
the first time even in four dimensions. For application
to a finite-temperature system, we set $D=4$ and replace
$L$ with $\beta=T^{-1}$ (where $T$ is the temperature of the system).
A peculiarity is known that $I(x)$ is linear in $x$ for small
$x$ when $D=4$. Because of this behaviour of $I(x)$, the free
energy (density) as a function of r has a minimum located
around $\phi\approx g^2/4\pi$. This suggests that a symmetry breaking
occurs in the YM system at high temperature \cite{11}.
Using the result here, it is revealed that the inclusion
of arbitrary numbers of fermions belonging to fundamental or adjoint representations of SU(2) does not
modify the vacumm expectation value of $r$ to the order
of $g^2$ \cite{13}.

After completion of the first draft of the present paper,
we find a very recent paper of \cite{18}. In the paper, Belyaev
showed that the vacuum condensation $\phi\approx g^2/4\pi$ is not
a true order parameter and the study of Wilson loop
exhibits that there is no symmetry breaking in finite-
temperature pure YM systems. Invariance of the
minimum at $\phi\approx g^2/4\pi$ when fermions are added must
be crucial in the analysis of Wilson loop in the system
including fermions. We will treat the subject elsewhere.

\section{Beyond the two loop}
As in the previous section, one can calculate the vacuum
energy to higher-order in g by computing the higher-loop
diagrams (whereas, for instance in four dimensions, we
should be careful to sum over loops which yields the
contribution of the order $g^3$ \cite{10}). If the coupling $g$ takes
a large value, however, such a perturbative approach
loses its reliability. When we consider QCD as a non-Abelian gauge
theory, validity of perturbation is confirmed in certain occasions; on
the other hand it is necessary to examine non-perturbative effects
quantitatively as well as qualitatively.

As an example, we take the gauge field condensation
mentioned in the previous section. The minimum of free
energy has been obtained for small $g$. For large $g$, this
perturbative analysis is no longer reliable. However, we
are interested in another possibility of gauge field
condensation in the non-perturbative region.

In QCD, it is significant to study the coupling dependence of various physical quantities in order to investigate
not only cosmology of the early universe but also experiments of heavy nuclei.

It is also pointed out that transition of a certain model
of unified theory cries out for consideration of non-perturbative
effect in gauge theory \cite{14}. Effort in treating the
non-perturbative phenomena analytically is of importance.

From the point of view of the computational technique, it will be convenient to treat all-order effect by
an extension of one-loop treatment. In this paper we
try to take out much information of non-perturbative
physics from the representation of vacuum energy.

In Sect. 2, we have calculated the two-loop vacuum
diagrams. The D-dependence of the coefficient of each
diagram is as follows (see (\ref{2.2a}, \ref{2.2b}, \ref{2.2c}) and
Fig.~1): for (a) $\approx D^2$, for (b) $\approx D^1$, and for (c)
$\approx D^0$ for large
$D$. Namely, the graph including YM four-point interaction dominates
for large $D$. Here $D$ comes from the trace of the metric
at a closed loop. The graph (b) is sub-dominant because
three-point interaction has a derivative coupling. The
contribution of the graph (c) does not have the trace of
the metric.

Therefore it is conceivable that for large $D$ only
four-point interaction is important and this simplifies the
treatment of YM interactions of higher order. In the rest
of this section we consider an approximation method
based on this observation.

The $D$-dependence appears also in $I(x)$, that is, in the
momentum integrations. Thus the mass insertian reduces
the order of $D$. As we can see later, this effect is naturally
involved by our approximation.

First of all, we write down the YM action where three-point
interactions are omitted: 
\begin{eqnarray}
\frac{1}{4}{\rm tr~}F^2&\approx&\frac{1}{4}\sum_{\mu\nu}\sum_{a=1}^3
(D^B_\mu a^a_\nu-D^B_\nu a^a_\mu)^2+(\mbox{gauge-fixing term})
\nonumber \\
& &+g^2\{(a^2_\mu a^3_\nu-a^2_\nu a^3_\mu)^2+
(a^3_\mu a^1_\nu-a^3_\nu a^1_\mu)^2+
(a^1_\mu a^2_\nu-a^1_\nu a^2_\mu)^2\}\,,
\label{3.1}
\end{eqnarray}
where $a^a_\mu$ is the quantum fluctuation of the gauge field while
$D^B_\mu$ denotes the covariant derivative involving the background
classical fields. If we further neglect the interactions which take the
form of $a^1_\mu a^1_\nu a^2_\mu a^2_\nu$, we obtain
\begin{eqnarray}
& &\sum_{\mu\nu}\left[\frac{1}{2}\sum_a(D^B_\mu
a_\nu^a)^2+\frac{1}{2}g^2\{(a^2_\mu a^2_\mu)(a^3_\nu
a^3_\nu)\right.\nonumber
\\ & &\left.+(a^3_\mu a^3_\mu)(a^1_\nu
a^1_\nu)+(a^1_\mu a^1_\mu)(a^2_\nu
a^2_\nu)\right]\,.
\label{3.2}
\end{eqnarray}
The reason why the terms like $a^1\mu a^1_\nu a^2_\mu a^2_\nu$ can be
omitted is that we cannot make the graph like Fig.~3 by only
use
of those terms, while graphs of the type of Fig.~4 can be
made: although both contributions of Figs.~3 and 4 are
of the order of $g^4$, the contribution of Fig.~3 is superior
to that of Fig.~4 by factor of the order of $D$ which comes
from trace of the metric associated with each loop.
Therefore as long as we regard $D$ as a large number, the
YM Lagrangian is well approximated by (\ref{3.2}).

\begin{figure}[ht]
\begin{center}
\includegraphics[width=2cm]{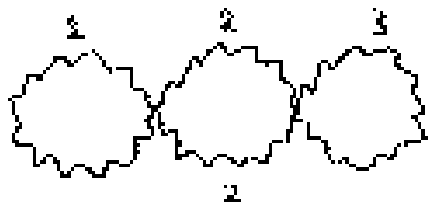}
\caption{The graph of the order of $g^4$ which can be led from the
effective Lagrangian (\ref{3.2})
}
\label{f3}
\end{center}
\end{figure}
\begin{figure}[ht]
\begin{center}
\includegraphics[width=2cm]{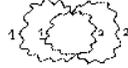}
\caption{The graph of the order of $g^4$ constructed by the interaction
terms like $a^1\mu a^1_\nu a^2_\mu a^2_\nu$ which are not involved in
the effective Lagrangian (\ref{3.2})
}
\label{f4}
\end{center}
\end{figure}

Owing to this simplification, the calculation of the
vacuum energy becomes very transparent. We can utilize
auxiliary fields \cite{15} to treat the non-linear interaction in
the Lagrangian (\ref{3.2}).

Using the knowledge that the matrix
\begin{equation}
\left(\begin{array}{ccc}
0 & 1 & 1\\
1 & 0 & 1\\
1 & 1 & 0
\end{array}
\right)\,,
\label{3.3}
\end{equation}
has its inverse matrix 
\begin{equation}
\frac{1}{2}\left(\begin{array}{rrr}
-1 & 1 & 1\\
1 & -1 & 1\\
1 & 1 & -1
\end{array}
\right)\,,
\label{3.4}
\end{equation}
we rewrite the Lagrangian as
\begin{eqnarray}
& &\frac{1}{2}\sum_{\mu\nu}\sum_a(D^B_\mu a^a_\nu)^2+\frac{1}{2}
\sum_\mu\sum_a \chi^a a^a_\mu a^a_\mu+\frac{1}{8g^2}\nonumber \\
& &\cdot[(\chi^1)^2+(\chi^2)^2+(\chi^3)^2-2(
\chi^2\chi^3+\chi^3\chi^1+\chi^1\chi^2)]\,.
\label{3.5}
\end{eqnarray}
The use of the auxiliary fields $\chi^a$ ($a = 1, 2, 3$) enable us to
rewirte the action in the bilinear form of $a_\mu$'s.

We calculate the vacuum energy in the presence of
the background field (\ref{2.1}). The background fields and
metric are assumed to be the same as in the previous
section. Because of the symmetry, we can set $\chi^1=\chi^2$ and
then the formal expression of the one-loop vacuum energy
including the auxiliary fields are given as
\begin{eqnarray}
& &\frac{1}{8g^2}((\chi^3)^2-4\chi^1\chi^3)\nonumber \\
& &+D^*\int \frac{d^dp}{(2\pi)^d}\sum_{k=-\infty}^\infty
\ln\left\{p^2+\left(\frac{2\pi k+\phi}{L}\right)^2+\chi^1\right\}
\nonumber \\
& &+\frac{D^*}{2}\int \frac{d^dp}{(2\pi)^d}\sum_{k=-\infty}^\infty
\ln\left\{p^2+\left(\frac{2\pi k}{L}\right)^2+\chi^3\right\}\,.
\label{3.6}
\end{eqnarray}

Since we are ignoring the contribution of ghost fields,
we denote the coefficient of the integrals using the
constant $D^*$ which is the same order as $D$. Later we will
``adjust'' the constant $D^*$.

After regularization of (\ref{3.6}), we obtain the following
expression in the large $D$ (also $D^*$) limit:
\begin{eqnarray}
& &\frac{1}{8g^2}((\chi^3)^2-4\chi^1\chi^3)\nonumber \\
&
&-4D^*\left[\sum_{k=1}^\infty\left(\frac{\sqrt{\chi^1}}{2\pi
kL}\right)^{D/2}K_{D/2}(\sqrt{\chi^1}Lk)\cos k\phi\right]
\nonumber \\
& &-2D^*\left[\sum_{k=1}^\infty\left(\frac{\sqrt{\chi^3}}{2\pi
kL}\right)^{D/2}K_{D/2}(\sqrt{\chi^3}Lk)\right]\,,
\label{3.7}
\end{eqnarray}
where $K_n(x)$ is the modified Bessel function. Eliminating
the auxiliary fields by applying the equations of motion
for $\chi$, we obtain the vacuum energy including the
(four-point) interaction effect. For larger $D$, more exact
result for energy is expected to be given.

As a check, we will show that the result of the one- and
two-loop order is obtained when $g^2 \ll 1$.

For small $g$, the equations of motion for $\chi$ obtained
from (\ref{3.7}) can be solved approximately as
\begin{eqnarray}
\chi^1&=&g^2D^*\frac{\Gamma\left(\frac{D}{2}-1\right)}{2\pi}
\left(\frac{1}{\pi L^2}\right)^{D/2-1}\sum_{k=1}^\infty\frac{1+\cos
k\phi}{k^{D-2}}\,,\label{3.8a}\\
\chi^3&=&g^2D^*\frac{\Gamma\left(\frac{D}{2}-1\right)}{2\pi}
\left(\frac{1}{\pi L^2}\right)^{D/2-1}\sum_{k=1}^\infty\frac{2\cos
k\phi}{k^{D-2}}\,,
\label{3.8b}
\end{eqnarray}
Thus the vacuum energy for YM system in $R^d\times S^1$ for
small coupling $g$ is obtained to the order of $g^2$ as:
\begin{eqnarray}
& &(\mbox{vac. energy of YM for small } g^2)
\nonumber \\
& &\approx -D^*\frac{\Gamma\left(\frac{D}{2}\right)}{\pi^{D/2}L^D}
\sum_{k=1}^\infty\frac{1+\cos
k\phi}{k^{D}}\nonumber \\
&
&+\frac{g^2}{2}\left\{D^*\frac{\Gamma
\left(\frac{D}{2}-1\right)}{2\pi^{D/2}L^{D-2}}
\right\}^2
\left(\sum_{k=1}^\infty\frac{\cos
k\phi}{k^{D-2}}\right)
\left(\sum_{k=1}^\infty\frac{2+\cos
k\phi}{k^{D-2}}\right)\,.
\label{3.9}
\end{eqnarray}
Therefore we find that the substitution $D^*=D-2$ leads
to the exact result of perturbative calculation of the
two-loop diagrams (see (\ref{2.4})).

In physical meaning, $\chi$ stands for the ``mass'' of the
gauge bosons. In finite-temperature systems, this is finite
in general \cite{10}. In four dimensions ($D=4$), let us investigate
the next-order contribution in $g^2$. We rewrite $T=L^{-1}$ and 
$\langle\phi\rangle$ is
assumed to vanish. Using asymptotic expansion of $K_1(x)$, we obtain
($\chi^1=\chi^3=\chi$)
\begin{equation}
\frac{\chi}{T^2}=\frac{g^2D^*}{6}-\frac{1}{2\sqrt{6}}(g^2D^*)^{3/2}
+\cdots\,.
\label{3.10}
\end{equation}
Though the result is not exactly coincident with the
known result \cite{10}, we obtain the correct order of
magnitude of coefficients and the correct fractional
dependence on $g$ in the next-leading term \cite{10}. Our
approximation is not so bad even in four dimensions,
$D=4$.

Another consequence which can be compared with
the perturbation results is the dependence on gauge
parameter. So far we take Feynman gauge, $\xi=1$; to take
other choice is straightforward because we only need the
technique of the one-loop calculation. It is known that
there is in fact gauge-dependence of $(\xi-1)^1$ at the two-
loop order, while a naive counting suggests at most
$(\xi-1)^3$ dependence \cite{16}. By our method we find $(\xi-1)^1$
dependence in the order of $g^2$, though we do not give the
explicit calculation here.

Now we consider some examples for applications of
our method.

If we consider $S^1$ as an extra space, we can calculate
vacuum energies for YM fields in the Kaluza-Klein
background. The one-loop effects of various fields have
been investigated in many authors \cite{5,7,8}. The approximation
scheme in the present paper is effective in Kaluza-Klein theories
because the number of dimensions may be arbitrarily large. Explicit
calculations will be shown in \cite{19}.

Next we consider finite-temperature systems. We set
$T=L^{-1}$, where $T$ is the temperature of the system, and
$D=4$. At first we assume the background field $\langle\phi\rangle= 0$.
The free energy of free massless particle system is proportional to 
$T^4$. The effect of YM interaction modifies the relation. To see this,
we define an ``effective degrees of freedom'', denoted by $|F|$, as
\begin{equation}
(\mbox{effective degrees of freedom})=\frac{|\mbox{free energy
density}|}{T^4}\,. 
\label{3.11}
\end{equation}
By numerical calculation, we can obtain free energy of
YM boson gas. The result is shown in Fig.~5. It is
physically natural to see the effective degrees of freedom
decreases when $g^2$ becomes large. Since its decrease is
very moderate, the change on the temperature through
the running of the coupling $g$ is expected to be small.
Therefore, for example, the analysis of the finite-volume
effect \cite{17} is important on the physical confinement.

\begin{figure}[ht]
\begin{center}
\includegraphics[width=5cm]{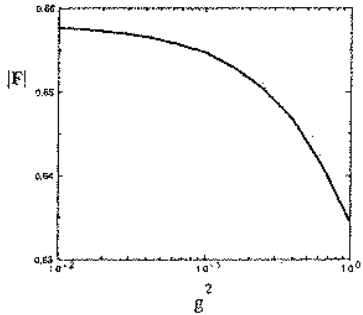}
\caption{The plot of the effective degrees of freedom vs. $g^2$
}
\label{f5}
\end{center}
\end{figure}

We can calculate the vacuum energy even in the
presence of the background gauge field. In higher
dimensional theory, the symmetry breaking by the field
is known as Hosotani mechanism \cite{7}. At the one-loop
level, the mechanism is investigated by many authors.
We will report the application of our method in higher
dimensions elsewhere \cite{19}.

\begin{figure}[ht]
\begin{center}
\includegraphics[width=5cm]{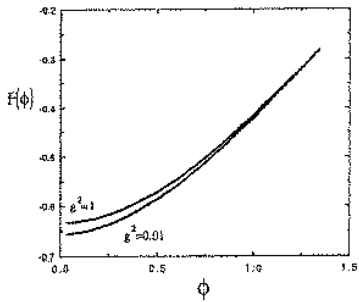}
\caption{The plot of 
$F(\phi)=(\mbox{the free energy density}/T^4)$ vs. $\phi$ for 
$g^2=0.01$ and $g^2=1$
}
\label{f6}
\end{center}
\end{figure}

In four dimensions, the background field at finite
temperature is also useful to analyze the finite-volume
effect in YM system \cite{17}. In Fig.~6, 
$F(\phi)=(\mbox{the free energy})/(\mbox{temperature})^4$ is plotted
against $\langle\phi\rangle$. (Note: $|F|=F(0)$.) The minimum remains
located at $\langle\phi\rangle=0$. Unfortunately, non-perturbative
minima have not been found by the present analysis.

\begin{figure}[ht]
\begin{center}
\includegraphics[width=5cm]{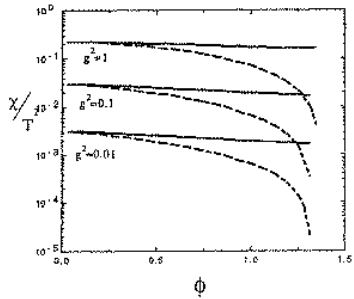}
\caption{The plot of the dimensionless combination $\chi^a/T^2$ vs.
$\phi$
for $g^2=0.01, 0.1$ and $1$. The solid lines stand for $\chi^3/T^2$,
while the broken lines stand for $\chi^1/T^2$
}
\label{f7}
\end{center}
\end{figure}

The values of $\chi^a/T^2$ in the presence of $\langle\phi\rangle$ are
given in Fig.~7.

The curvature of the change in $\phi$ near zero becomes
smaller for larger $g$; thus the finite-volume effect, which
is obtained by the integration with respect to $\phi$ around
$\langle\phi\rangle=0$ \cite{17}, is expected to increase slightly if
the coupling $g$ becomes large.

We hope to study more generic cases, such as with
$SU(3)$ YM fields, and thorough analyses on the various
systems in the near future.

\section{Summary and future problems}
We have obtained the vacuum energy for YM system in
the D-dimensional space $S^1\times R^d$ $(D=d+1)$. We have
taken the presence of the background gauge field into
consideration. In the present paper we have performed
the two-loop calculation in $D$ dimensions. We have
shown the approximation scheme which is suitable for
large $D$ and by the use of this one can investigate the
effects of YM interaction. For small $g$, the YM
self-coupling, we have shown the reconstruction of free
energy to the two-loop order (i.e., the functional form
and the dependence on $D$) by our method.

Our computational method is especially well-suited
for Kaluza-Klein theories in higher dimensions. Even if
the extra space has a complicated structure, we can study
the higher-loop effects by our method similarly to the
one-loop technique. The study of this is currently in
progress.

Similar analyses of higher-loop effects on the vertex
and the propagator graphs are possible, though we have
treated only the vacuum graph in this paper. This subject
is an interesting one which we wish to study.

Another important task is the extension to large
symmetry groups. Besides the application to QCD, it is
claimed that the behavior of the free energy for the system
of gauge bosons which interact themselves strongly is
important in the symmetry breaking in the early universe.

We want to develop the method which is more effective
in various situations.

We hope that various numerical calculations which
can be treated by our method will be reported in near
future.

\section*{Acknowledgments} The authors would like to thank A. Nakamula
and T. Koikawa for some useful comments. One of the authors
(KS) thanks A. Nakamichi for helpful advice. He would also like
to thank A. Sugamoto for reading this manuscript. KS is indebted
to Soryuusi shogakukai for financial support. He also would like
to acknowledge financial aid of Iwanami F\=ujukai.


\end{document}